\documentclass[]{aa}

\usepackage{amsmath}
\usepackage{amssymb}
\usepackage{url}
\usepackage{graphicx}
\usepackage{times}
\usepackage[utf8]{inputenc}
\usepackage{xspace}
\usepackage{color}
\usepackage{multirow}
\usepackage{multicol}
\usepackage{float}
\usepackage{verbatim}
\usepackage{mathtools, cuted}
\usepackage[normalem]{ulem}
\usepackage{soul}
\usepackage{booktabs}
\usepackage[dvipsnames]{xcolor}

\bibpunct{(}{)}{;}{a}{}{,}

\usepackage{etoolbox}
\makeatletter
\patchcmd\@combinedblfloats{\box\@outputbox}{\unvbox\@outputbox}{}{%
  \errmessage{\noexpand\@combinedblfloats could not be patched}%
}%
\makeatother

\def\apjl{ApJL}
\def\apjs{ApJSS}

\def\aap{A\&A}

\def\tv{\theta_\mathrm{v}}

\def\eg{e.g.\xspace}

\def\msun{\,\rm M_\odot}

\begin{document}

\title{Light-curve models of black hole - neutron star mergers: steps towards a multi-messenger parameter estimation\thanks{Light-curves in Fig. \ref{fig:example_lightcurves} are available in electronic form
at the CDS via anonymous ftp to \url{cdsarc.u-strasbg.fr} (130.79.128.5)
or via \url{http://cdsweb.u-strasbg.fr/cgi-bin/qcat?J/A+A/}}}

\titlerunning{BHNS merger EM counterparts}
\author{C.~Barbieri\inst{\ref{unimib},\ref{infn.mib}\thanks{E--mail: c.barbieri@campus.unimib.it}}, O.~S.~Salafia\inst{\ref{oab.me},\ref{infn.mib}}, A.~Perego\inst{\ref{infn.mib},\ref{unitn}}, M.~Colpi\inst{\ref{unimib},\ref{infn.mib}} and G.~Ghirlanda\inst{\ref{oab.me},\ref{unimib}}}

\institute{Università degli Studi di Milano-Bicocca, Dip. di Fisica ``G. Occhialini'', Piazza della Scienza 3, I-20126 Milano, Italy\label{unimib}\and INAF -- Osservatorio Astronomico di Brera, via E. Bianchi 46, I-23807 Merate, Italy\label{oab.me}  \and INFN -- Sezione di Milano-Bicocca, Piazza della Scienza 3, I-20126 Milano, Italy\label{infn.mib} \and Universit\`a degli Studi di Trento, Dip. di Fisica, via Sommarive 14, I-38123 Trento, Italy \label{unitn}}

\authorrunning{C.~Barbieri  et. al}

\date{Received xxx / Accepted: xxx}

\abstract{
In the new era of gravitational wave (GW) and multi-messenger astrophysics, the detection of a GW signal from the coalescence of a black hole - neutron star (BHNS) binary remains a highly anticipated discovery. This system is expected to be within reach of the second generation of ground-based detectors. In this context, we develop a series of versatile semi-analytical models to predict the properties of all the electromagnetic (EM) counterparts of BHNS mergers. We include the nuclear-decay-powered kilonova emission, its radio remnant, the prompt emission from the jet, and the related afterglow. The properties of these counterparts depend upon those of the outflows that result from the partial disruption of the NS during the merger and from the accretion disc around the remnant, which are necessary ingredients for transient EM emission to accompany the GW signal. We therefore define ways to relate the properties of these outflows to those of the progenitor binary, establishing a link between the binary parameters and the counterpart properties. From the resulting model, we anticipate the variety of light curves that can emerge after a BHNS coalescence from the radio up to gamma-rays. These light curves feature universal traits that are the imprint of the dynamics of the emitting outflows, but at the same time, they show a clear dependence on the BH mass and spin, but with a high degree of degeneracy. The latter can be deduced by a joint GW - EM analysis. In this paper, we perform a proof-of-concept multi-messenger parameter estimation of a BHNS merger with an associated kilonova to determine how the information from the EM counterpart can complement that from the GW signal. Our results indicate that the observation and modelling of the kilonova can help to break the degeneracies in the GW parameter space, leading to better constraints on the BH spin, for example.

}

\maketitle

\section{Introduction}
The global network of advanced gravitational wave (GW) detectors, currently consisting of the two Advanced LIGO in the United States  and Advanced Virgo in Italy, is constantly improving in sensitivity, and the new detector KAGRA, located in Japan, is due to enter the network soon \citep{ranges,Aasi2015,Acernese2015,Aso2013}. The capabilities of the network have been demonstrated by a spectacular sequence of detections of binary black hole inspirals and mergers since September 2015 \citep{Abbott18-10-bh}. During the final month of the latest observing run, the network also detected the first GW signal from the inspiral of a double neutron star (NS) binary \citep{GW170817}. We are therefore in a position to expect the first detection of a stellar-mass black hole - neutron star binary (BHNS) to take place in the near future \citep{Abadie2010}.
This new source of GWs is one of the most promising targets for multi-messenger astronomy.

The rate of BHNS coalescences from population synthesis models is expected to be between $10^{-9}$ and  $\sim 10^{-6}$ Mpc$^{-3}$  yr$^{-1}$ \citep{Abadie10,Clark2015,Dominik2015,Mapelli2018}.
Based on observations of black hole - black hole (BHBH) binary coalescences, \cite{Abbott-rates2018} inferred a rate in the interval between $10^{-8}$ and $3\times 10^{-7}$ Mpc$^{-3}$ yr$^{-1}$, which is comparable.
On the other hand, the non-detection of a BHNS event during the LIGO O1 science run allowed to place a 90\% upper limit on the BHNS coalescence rate of $3.6 \times 10^{-6}$ Mpc$^{-3}$  yr$^{-1}$, assuming $5\msun$ and $1.4\msun$ for the BH and NS mass, respectively, and an isotropic spin distribution \citep{Abbott-rates2018}, with slightly more constraining values for higher BH masses. Unless the actual rate turns out to lie at the low end of the current estimates, we can reasonably expect the first detection of GW from this class of sources to take place in the near future, possibly during the upcoming O3 observing run.

Mergers of BHNS are exquisite probes of gravity and nuclear matter under extreme conditions, and are expected to display a rich variety of signals as they are likely to encompass a larger interval of masses and mass ratios than binary neutron star (NSNS) mergers. The known NSs, members of galactic binaries, have masses between $\sim 1.2\msun$ and $\sim 2\msun$ \citep{Ozel2012,Ozel2016}, and 
the masses of the two coalescing NSs in GW170817 fall in the same interval \citep{GW170817}. As demonstrated by the recent ground-breaking detections of BHBH mergers by the LIGO/Virgo Collaboration \citep{GW150914,GW151226,GW170608,GW170814},
the BH mass interval is significantly wider than that inferred from the observations of galactic X-ray binaries \citep{Ozel2010}, now ranging between $7.6^{+1.3}_{-2.1}$ and $50.6^{+16.6}_{-10.2}$ for the ten discovered GW sources \citep{Abbott18-10-bh}.

Fully general-relativistic (GR) numerical simulations of BHNS mergers show that when a coalescence is imminent, the NS is either torn apart (partially or totally) by the BH tidal field outside the innermost stable circular orbit (ISCO) or is swallowed directly by the BH. The NS fate, in a BHNS binary merger, depends on the mass ratio of the two compact objects, on the spins, and on the NS tidal deformability \citep{Shibata11, Foucart2012,Kyutoku2015,Kawaguchi2015,Foucart2018}. Higher BH spins and lower BH and NS masses set the most favourable conditions for the disruption of the star, with the GW signal carrying valuable information on the mass ratio, BH spin, and NS equation of state (EoS) \citep{Bildsten1992,Shibata2009,Foucart2013a,Foucart2013b,Kawaguchi2015,Pannerale2015a,Pannerale15b,Hinderer2016,Kumar2017}. 
Further exquisite information on the rich physics that accompanies the merger can be inferred from the electromagnetic (EM) transients that are expected to follow the disruption of the NS.
In this case, neutron-rich debris remains outside the BH innermost stable circular orbit (ISCO) in the form of a neutrino-cooled accretion disc and of a variable amount of dynamical ejecta \citep{DiMatteo2002,Chen2007,Shibata11,Foucart2012,Janiuk2013,Kawaguchi2015}. 
 
The detection of the first GW signal from the double neutron star (NSNS) binary GW170817 \citep{GW170817} and the discovery of its EM counterparts \citep{gw170817em} confirmed earlier predictions (e.g.~\citealt{Eichler1989,Narayan1992}) that NSNS mergers are one viable progenitor of short-duration gamma-ray bursts (SGRBs) and are the production sites of $r$-process elements that power the kilonova (KN) emission \citep{Lattimer1974,Li1998,Metzger2017}. Owing to the possibly high mass that is dispersed during the merger, BHNS coalescences might also give rise to both a GRB and a KN  emission \citep{Ascenzi2018,Ascenzi2019}. Recent GR magnetohydrodynamics simulations \citep{Shapiro-jet2015,Shapiro2017,Pasch2017,Ruiz-Jet-2018} indeed indicate that the remnant of a magnetized BHNS merger can launch a jet, possibly powering an SGRB. 

Several authors started to explore the properties of KN emission in BHNS coalescences, mainly through radiation-transfer simulations in a restricted interval of the BHNS merger parameter space \citep{Tanaka13,Tanaka2014,Fernandez2017}. Interestingly, these authors find that the radioactively powered emission from these binaries can be more luminous than that from NSNS mergers because more mass is ejected from the NS disruption. 
\cite{Kawaguchi2016} explored KN emission from a wider range of BHNS mergers using a semi-analytical model and fitting formulae for the mass and velocity of the dynamical ejecta that were calibrated using a larger set of GR numerical simulations of BHNS coalescences by the Kyoto group. 
 
The population of SGRB progenitors has been studied for years, but only within the limited view of their prompt emission (e.g. \citealt{Ghirlanda2016}), and of their afterglow emission when present \citep{Fong2015}. With the opening of the new GW era, we have the unique opportunity to explore for the first time the multi-messenger outcome of mergers of compact objects in all their flavours, comprising BHNS coalescences, and to discern whether SGRBs have only one progenitor, that is, NSNS mergers, or whether BHNS coalescences in a certain range of masses, mass ratios, and BH spins can also power the transient emission observed in the SGRB population.

We here build a series of models to predict the expected multi-wavelength emission that accompanies BHNS mergers.  We include most of the jet- and KN-related EM components: the nuclear-decay-powered KN emission (both from dynamical ejecta and disc winds), its radio remnant (KNR), the prompt emission from the jet, and the related afterglow.

Prospects for multi-messenger analysis for BHNS mergers have been discussed in \cite{Pannarale2014}. \cite{Coughlin2017,Coughlin2018} and \cite {Coughlin2019} showed that a combined analysis of EM and GW data from an NSNS or BHNS merger helps to constrain the intrinsic parameters of the binary and the equation of state (EoS) of matter at supra-nuclear densities. \cite{Hinderer2018} presented an example of a multi-messenger parameter estimation for a BHNS merger, under the assumption that GW170817 was a binary of this type. Much in this spirit, we aim not only at anticipating the properties of the EM counterparts of BHNS coalescences, but also at setting a framework for joint GW and EM analysis. 

The paper is organised as follows. In \S\ref{outflows} we briefly present the different outflows from a BHNS merger producing the EM emission. In \S \ref{parameters} we introduce the set of parameters describing BHNS binaries. In \S\ref{merger_result} we explain how we model the masses of the  accretion disc and dynamical ejecta, respectively. \S\ref{Kilonova} illustrates the model for the KN emission, while in \S\ref{knaft} we describe the kilonova radio remnant (KNR). In \S\ref{jet} we present the model for the launch of a relativistic jet. Its associated emissions (GRB prompt and afterglow) are described in \S\ref{prompt} and \S\ref{GRBafterglow}, respectively. Example light curves and their dependence on the BH parameters are presented in \S\ref{lightcurves}. In \S\ref{testcase} we present how the BH spin can be constrained with the observation of an EM counterpart. Finally, we list in Appendix A the constraints on the BH and NS masses obtained from GW analysis.

Throughout this work we assume a $\Lambda$CDM cosmology with parameters $\Omega_\textrm{M}=0.3065$, $\Omega_\lambda=0.6935$, $\Omega_\textrm{k}=0.005$, and $h=0.679$ as estimated by \cite{Planck}.

\section{Outflows from a BHNS merger}\label{outflows}

\begin{figure*}
    \centering
    \includegraphics[width=\textwidth]{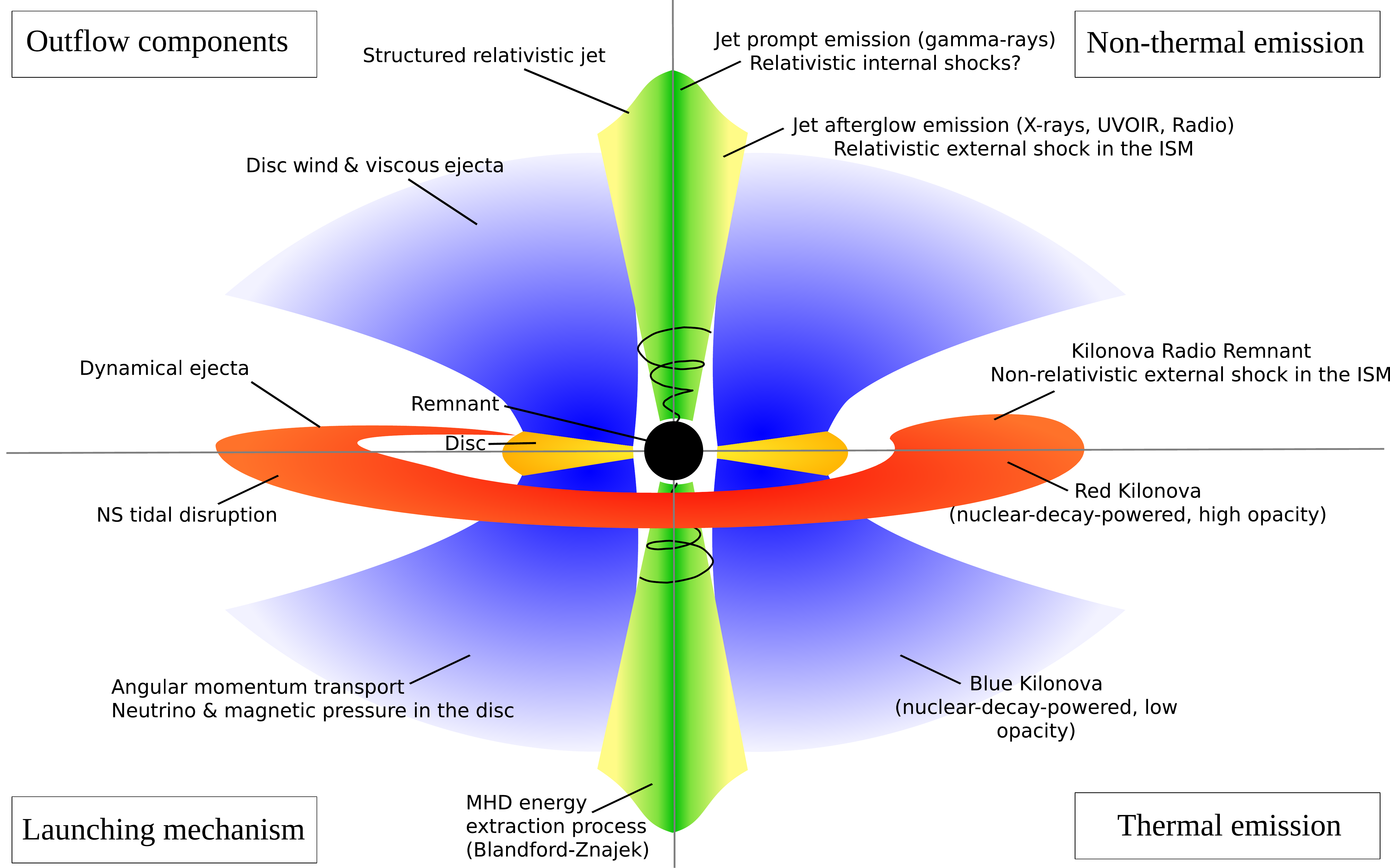}
    \caption{Electromagnetic counterparts we consider in this work. Here we sketch the merger remnant, a spinning black hole with an accretion disc, that is surrounded by various types of outflows. The sketch is divided into four quadrants. The upper left quadrant enumerates the outflow components we take into account. The lower left quadrant states the launching mechanism for each outflow. In the upper right quadrant we list the non-thermal emission components that arise from these outflows, either internally or upon interaction with the interstellar medium. In the lower right quadrant we list the thermal emission component, i.e. the red and blue KN emission from the dynamical and disc ejecta, respectively. }
    \label{fig:em_counterparts}
\end{figure*}

We focus on electromagnetic emission that arises from three types of outflows:
\begin{itemize}
    \item the dynamical ejecta, that is, the unbound material that results from the tidal disruption of the NS;
    \item the disc ejecta, that is,~the outflows that originate in the accretion disc (see \S\ref{Kilonova});
    \item the relativistic jet that may be launched by the remnant, powered by accretion and rotation.
\end{itemize}
The various outflows are illustrated in Figure~\ref{fig:em_counterparts}, along with brief descriptions of their launching mechanisms and of the electromagnetic emission components that arise either within the outflows themselves or upon their interaction with the interstellar medium (ISM). This figure is meant as a visual reference for the processes and phenomena described in this work.

\section{BHNS binary parameters}\label{parameters}
A BHNS binary is characterized by a number of intrinsic parameters: the BH and NS gravitational masses, $M_{\rm BH}$ and $M_{\rm NS}$, the effective tidal deformability of the system $\tilde{\Lambda}$, the component spins, $\mathbf{S}_{\rm BH}$ and $\mathbf{S}_{\rm NS}$, and $\iota_\textrm{tilt}$, the angle between the BH spin vector, and the orbital angular momentum vector. The parameter $\tilde{\Lambda}$ is a mass-weighted combination of the dimensionless quadrupolar tidal deformabilities of the binary components \citep{Raithel2018}. For a BHNS binary, because the BH is not deformable ($\Lambda_\textrm{BH}$=0), $\tilde{\Lambda}$ is defined as 
\begin{equation}
\tilde{\Lambda}=\frac{16}{13}\frac{(
M_\textrm{NS}+12
M_\textrm{BH})M_\textrm{NS}^4\Lambda_\textrm{NS}}{(M_\textrm{BH}+M_\textrm{NS})^5}.\end{equation}
The quantity $\Lambda_\textrm{NS}$ can be written as
\begin{equation}
\Lambda_\textrm{NS}=\frac{2}{3}k_2C_{\rm NS}^{-5}
,\end{equation}
where $C_{\rm NS}$ is the compactness $C_{\rm NS}=GM_{\rm NS}/(R_{\rm NS}c^2)$, with $R_{\rm NS}$ the NS radius, $G$ the gravitational constant, and  $c$ the speed of light. $k_2$ is the dimensionless tidal Love number $k_2=(3/2)G\lambda R_{\rm NS}^{-5}$ (\citealt{Flanagan2008}), where $\lambda$ is the quadrupolar polarisability, which represents the ratio of the induced quadrupole moment $Q_{ij}$ to the applied tidal field $E_{ij}$, namely $Q_{ij}=-\lambda E_{ij}$.

We calculated the NS compactness using the `C-Love' relation from \cite{Yagi2017}. In that work, the authors find an approximately universal (EoS-independent) relation between the NS compactness and the dimensionless tidal deformability, which takes the form
\begin{equation}\label{eq:Cns}
C_{\rm NS}=\sum_{k=0}^2a_k({\rm{ln}}\Lambda_{\rm NS})^k    
.\end{equation}
We used this formula with the best-fit coefficients $a_k$ as given in \cite{Yagi2017}. The modulus of either spin can be expressed in terms of the dimensionless spin parameter $\chi=c\vert {\bf S}\vert /(GM^2).$ 
We neglected the NS spin, that is,~we assumed $\chi_\mathrm{NS}\sim 0$, 
as BHNS are long-lived systems before they reach coalescence, and the NS (born rapidly spinning) had time to spin-down by dipole emission. Furthermore, the lack of matter accreting onto the NS prevents spin-up by recycling. Thus the NS spin before any tidal locking is expected to be negligible, and it remains small as the timescale for  tidal spin-up is much longer than the GW-driven inspiral time (\citealt{Kochanek1992} and \citealt{Bildsten1992}). 

The BH spin plays a key role in the dynamics of the merger. For simplicity, we here considered non-precessing binaries so that the BH spin vector can be either aligned ($\iota_\mathrm{ tilt} = 0^\circ$) or anti-aligned ($\iota_\mathrm{ tilt} = 180^\circ$) with the orbital angular momentum. Anti-aligned configurations were discarded as they favour the direct plunge of the NS, the BH having a larger ISCO, so that no debris remains to power an EM counterpart. Therefore we excluded binaries with anti-aligned spins and conservatively considered values of $\chi_{\rm BH}$ in the range $[0,1]$. 

The extrinsic parameters considered in our study are the luminosity distance $d_\textrm{L}$ and the viewing angle $\theta_{\rm view}$, that is,~the angle between the line of sight and the direction of the orbital angular momentum (we took $0^\circ\leq \theta_\mathrm{view}\leq90^\circ$, i.e.~we did not distinguish `face-on' from `face-off' sources here, as we assumed all outflows to be axisymmetric and to have identical properties above and below the orbital plane). Where not stated otherwise, we fixed the following values:
\begin{itemize}
\item{$M_{\rm NS}$}  . The masses of the two NSs in GW170817 ($1.46^{+0.12}_{-0.10}\msun$ and $1.27^{+0.09}_{-0.09}\msun$ respectively, \citealt{catalogo_GW}) fall within the distribution of NS masses in galactic binaries, and the values are close to those expected for newly born NS (\citealt{Ozel2016,GW170817}). In an isolated binary, the BH formation precedes that of the NS, being the relic of the heaviest star in the system, which evolves faster. As there is no mass exchange in the newly born BHNS binary, we conservatively adopted for the NS gravitational mass the value $M_\textrm{NS}=1.4\msun$.

\item{$\Lambda_\textrm{NS}$} . We assumed the NS dimensionless quadrupolar tidal deformability $\Lambda_\textrm{NS}=330$. This value is close to that predicted by the SFHo EoS \citep{SFHo} for a $1.4\msun$ NS, which is $\approx334$. This nuclear EoS is fully compatible with the present nuclear and astrophysical constraints, and it predicts a radius of $\sim$12 km for a NS of $\sim$1.4$\msun$, which is close to the central values for the NS radii deduced in the analysis of the GW signal from GW170817 \citep{LVC_EoS2017}.  
\item{$d_\textrm{L}$} . We assumed a luminosity distance $d_\textrm{L}=230$ Mpc, corresponding to a redshift of $z=0.054$. This value is representative of the anticipated BHNS detection range during the next LIGO/Virgo observing run O3 \citep{ranges};
\item{$\theta_{\rm view}$} . We assumed $\theta_{\mathrm {view}}=30^\circ$, which corresponds to the most likely inclination angle of GW-detected binaries assuming a homogeneous isotropic population of sources in Euclidean space-time \citep{Schutz2011}.
\end{itemize}

\noindent In what follows, we also need the NS baryonic mass
\begin{equation}\label{eq:Mb}
M_\textrm{b} = B.E. + M_\textrm{NS},\end{equation}
where $B.E.$ is the energy gained by assembling $N$ baryons.  
The binding energy can be expressed as a function of the NS mass and compactness through the simple relation \citep{Lattimer2001}
\begin{equation}
B.E. = M_\textrm{NS}\frac{0.6C_\textrm{NS}}{1-0.5C_\textrm{NS}}
\end{equation}
to yield
\begin{equation}
M_\textrm{b} = M_\textrm{NS}\Big(1+\frac{0.6C_\textrm{NS}}{1-0.5C_\textrm{NS}}\Big)    
.\end{equation}
For a $1.4\msun$ NS, we infer a compactness $C_\textrm{NS}=0.178$ and a baryonic mass $M_\textrm{b}=1.56\msun$ based on this relation. SFHo EoS predicts  $C_{\rm NS}\approx0.174$ and $M_{\rm b}\approx1.56\msun$, in good agreement with this estimate.

\section{Disc and ejecta mass}\label{merger_result}

Before the merger, the BH is described by its mass $M_{\rm BH}$  and spin $\chi_{\rm BH}$ which in turn 
determine the radius of the ISCO, $R_{\rm ISCO}.$ As the NS approaches the BH, the tidal forces increase. At a `tidal' distance $d_{\rm tidal}\sim (M_\mathrm{BH}/M_\mathrm{NS})^{1/3}R_\mathrm{NS}$, the gravitational acceleration due to the NS self-gravity equals the tidal acceleration by the BH.

If $d_{\rm tidal}<R_{\rm ISCO}$, the NS experiences a direct plunge, and little or no mass is left outside of the BH: in this case, no EM counterpart is expected.  Conversely, if $d_{\rm tidal}>R_{\rm ISCO}$ , the NS is effectively disrupted and the BH remnant is surrounded by matter, which is the condition for the production of the EM counterparts.

The total baryon mass $M_\mathrm{out}$ left outside the BH can be divided into two components: the disc, that is,~the bound material, and the dynamical ejecta, that is,~the unbound part. We indicate their masses as $M_\mathrm{disc}$ and $M_\mathrm{dyn}$, respectively.

When the BH mass is fixed, it is evident that the heavier the NS, the larger the minimum BH spin that is required to produce a significant amount of $M_\mathrm{out}$. The reason is that higher mass NSs are generally more compact, leading to a smaller $d_\mathrm{tidal}$, which in turn requires a smaller $R_{\rm ISCO}$ to unbind material, or in other words,~a larger BH spin. The same holds when the BH masses are increased and the NS mass is kept fixed: more massive BHs have larger gravitational radii, so that higher spins are needed to avoid a direct plunge.
Therefore, binaries with low mass ratios $Q=M_\mathrm{BH}/M_\mathrm{NS}$ and high BH spins $\chi_{\rm BH}$ provide the best parameter combination to maximise the baryon mass outside the BH and to produce an EM counterpart. 
 
For the same reasons, keeping the BH and NS masses fixed, we have that the smaller $\Lambda_\textrm{NS}$  (i.e.~the softer the EoS), the higher the BH spin that is required to produce the same amount of $M_\mathrm{out}$. In other words, a softer EoS leads to a more compact NS, which is more difficult to disrupt.

In order to compute the disc and ejecta masses and use them as input to the EM counterpart models, we parametrised them as a function of the BH and NS intrinsic parameters. We proceeded in a way similar to \cite{Salafia2017} (see also \citealt{Coughlin2017,Coughlin2018,Coughlin2019}). We computed $M_\mathrm{out}$ using the physically motivated formula from \cite{Foucart2018}. The free parameters of this formula have been calibrated based on a suite of numerical simulations of BHNS mergers. $M_\mathrm{out}$ depends on $M_\mathrm{BH}$, $M_\mathrm{NS}$, $M_\mathrm{b}$, $\chi_\mathrm{BH}$ , and $\Lambda_\textrm{NS}$. \cite{Kawaguchi2016} provided a similar formula for $M_\mathrm{dyn}$ (and for the ejecta rms velocity $v_\mathrm{dyn}$), which depends on $M_\mathrm{BH}$, $M_\mathrm{NS}$, $M_\mathrm{b}$, $\chi_\mathrm{BH}$, $C_\textrm{NS}$ , and $\iota_\mathrm{tilt}$. Here, for a given $\Lambda_\mathrm{NS}$, we computed $M_\mathrm{b}$ and $C_\mathrm{NS}$ by Eqs.~\ref{eq:Mb} and \ref{eq:Cns}, respectively.
Therefore, having the total mass remaining outside the BH and the mass of the ejecta, we obtain the disc mass by computing their difference:
\begin{equation}
M_{\rm disc}=\max\left[M_{\rm out}-M_{\rm dyn};0\right]
.\end{equation}

Figure~\ref{fig:masse} shows the parameter region where an accretion disc and/or dynamical ejecta are present after the merger.
In the $M_{\rm BH}-\chi_{\rm BH}$ parameter space, only the coloured regions correspond to binaries whose merger will produce an EM counterpart. It is apparent that low BH masses and high spins are required. 

For the values of $\Lambda_\textrm{NS}$, $M_\textrm{NS}$ , and $\iota_{\rm tilt}$ assumed in section \ref{parameters}, the maximum disc mass is $\approx0.4\msun$, while the maximum dynamical ejecta mass and velocity are $\approx0.1\msun$ and $\approx0.6~c$ , respectively.

\section{Kilonova}\label{Kilonova}
The decompression of cold NS matter in a BHNS merger has long been thought to be a possible site for the production of the heaviest elements in the Universe through $r$-process nucleosynthesis \citep{Lattimer1974}. This nucleosynthesis process takes place in the merger ejecta and proceeds far from the nuclear valley of stability. The radioactive decay of the freshly synthesised 
$r$-process nuclei powers the KN emission on a timescale ranging from a few hours to a few weeks after the merger.

Matter from a BHNS merger is expected to be ejected through different channels that are characterised by different ejection mechanisms, timescales, and matter properties. In this work, we consider three types of ejecta:
\begin{enumerate}
\item {Dynamical ejecta}, which is produced by tidal interactions on a timescale of a few milliseconds during the merger (\citealt{Kawaguchi2016}, \citealt{Radice2018_2}).
\item {Wind ejecta}, which is produced by an accretion disc through neutrino-matter interactions and magnetic pressure. This ejection mechanism takes place on a timescale of tens of milliseconds (\citealt{Ruffert1997}, \citealt{Dessart2009}, \citealt{Kiuchi2015}, \citealt{Fernandez2017}).
\item{Viscous ejecta}, which is also produced by the accretion disc through viscous processes of magnetic origin inside the disc. This ejection takes place throughout the duration of the accretion because it is related to angular momentum transport (\citealt{Fernandez2013}, \citealt{Radice2018}).
\end{enumerate}

In order to describe the KN emission from the wind and secular ejecta, we adopted the semi-analytical model described in \cite{Perego2017}. The model assumes axisymmetry along the rotational axis of the system and divides the polar angle $\theta$ into 30 slices, equally spaced in $\cos{\theta}$. Each component is characterised by a certain mass, $m_{\rm ej}$, an average radial expansion velocity, $v_{\rm ej}$, and an effective grey opacity $\kappa_{\rm ej}$  that may be dependent on $\theta$. The matter is assumed to expand homologously inside each slice. The matter distribution in velocity space \citep[based on numerical simulations, see][]{Rosswog2013} is described by $dm/dv \propto (1- \left( v/v_{\rm max} \right)^2)^3$, where $v_{\rm max}$ is the maximum ejecta velocity. The maximum and mean velocity are related by 
$v_{\rm max} = 128/35 \, v_{\rm ej}$. Inside each slice, thermal emission at the photospheric radius is computed according to the model presented in \cite{Grossman2014} and \cite{Martin2015}.

\begin{figure}
    \centering
    \includegraphics[width=\columnwidth]{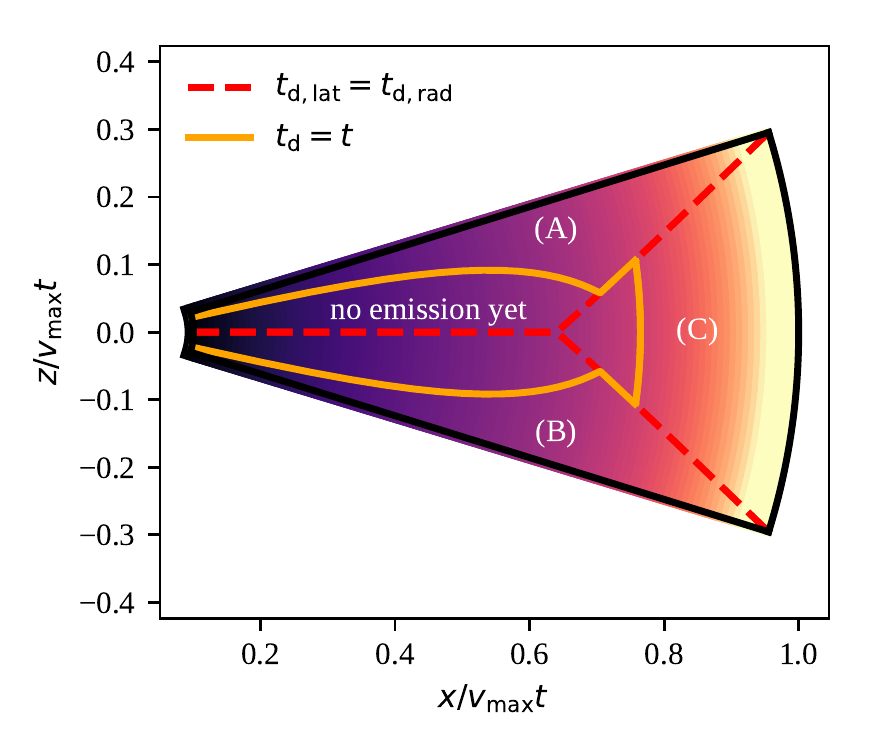}
    \caption{Sketch of the dynamical ejecta divided into different regions. The black line represents the edge of a section of the dynamical ejecta, assumed to have a crescent-like shape (see text). The inner colouring qualitatively depicts the decrease of density outwards. The $z=0$ line represents the equatorial plane. The red dashed line divides the ejecta into three regions, based on the edge to which the diffusion time is shortest. This is the upper latitudinal edge for region A, the lower edge for region B, and the radial edge for region C. The orange solid line separates the part of the ejecta for which radiation can diffuse to the relevant edge -- and where emission is thus possible -- from the part for which radiative diffusion has not yet reached the surface. 
    }
    \label{fig:dyn_ej_emission_regions}
\end{figure}

The dynamical ejecta in BHNS mergers are not axisymmetric along the rotational axis of the system. As shown in~\cite{Kawaguchi2016} and \cite{Fernandez2017}, for instance, the typical dynamical ejecta geometry is a crescent that is located close to the equatorial plane and extends azimuthally over approximately half of the equatorial plane (the azimuthal aperture is $\phi_\mathrm{dyn}\sim\pi$ rad), and latitudinally over an angle $\theta_\mathrm{dyn}\approx0.2-0.5$ rad. While an analytical model for the emission from these ejecta has been described in \cite{Kawaguchi2016}, it assumes a uniform velocity distribution, and it relies on rescaling of the spectrum from a single radiative transfer simulation from \cite{Tanaka2014}. An extension of this model, which accounts for both an inhomogeneous distribution of mass in the latitudinal direction and for a radial velocity profile, was recently presented in \cite{Huang2018}. We find the approximations used by the authors in treating the photon diffusion depth unsatisfactory, however. We therefore seek here to define a more general, while still simple, model. We assumed the same mass distribution in velocity space as the other ejecta. We used as Lagrangian coordinate for the dynamical ejecta the velocity $v$. We refer to the part that moves at a given velocity as a `shell'. Each shell emits from its latitudinal edge and contributes to the emission in the radial direction. We employed a simple diffusion approximation to compute both emissions, similar to the approximation employed in \cite{Grossman2014}.
We took $\theta$ to be the angle from the equatorial plane, so that the ejecta extend from $\theta=-\theta_\mathrm{dyn}$ to $\theta=\theta_\mathrm{dyn}$, and we focused on the upper half of the ejecta (positive $\theta$), as the same arguments with reversed signs hold for the lower part. The latitudinal diffusion time for radiation produced at an angle $\theta$ diffusing upwards in the shell is given by
\begin{equation}
t_{\mathrm{d,lat}}\sim{\frac{(\theta_\mathrm{dyn}-\theta)^2\kappa_\mathrm{dyn}\, dm/dv}{c\,\theta_\mathrm{dyn}\phi_\textrm{dyn}\,t}}
,\end{equation}
where $\kappa_\mathrm{dyn}$ is the effective grey opacity of the dynamical ejecta. 
The diffusion time from the shell to the surface in the radial direction is instead
\begin{equation}
t_{\mathrm{d,rad}}\sim{\frac{\kappa_\mathrm{dyn} m_\mathrm{dyn}(>v)(v_\mathrm{max}-v)}{c\,\theta_\mathrm{dyn}\phi_\mathrm{dyn} v^2 t}},   
\end{equation}
where $m_\mathrm{dyn}(>v)$ is the mass in the ejecta with velocity higher than $v$. We can thus find the angle $\theta_\mathrm{lat}(v)$ above which the diffusion time to the latitudinal surface is shorter than the time to the radial surface, which is given by
\begin{equation}
    \theta_\mathrm{lat}(v) = \theta_\mathrm{dyn}- \min\left(\theta_\mathrm{dyn},\sqrt{\frac{m_\mathrm{dyn}(>v)(v_\mathrm{max}-v)}{v^2 dm/dv}}\right)
.\end{equation}
We assumed that this angle divides the ejecta into three parts, each emitting only in the direction of the shortest diffusion time, as shown by the red dashed line in Figure~\ref{fig:dyn_ej_emission_regions}. For region A in the figure, the diffusion time equals the elapsed time at an angle $\theta_\mathrm{d}(v,t)$ given by
\begin{equation}
    \theta_\mathrm{d}(v,t) = \theta_\mathrm{dyn}- t\sqrt{\frac{c\,\theta_\mathrm{dyn}\phi_\mathrm{dyn}}{\kappa_\mathrm{dyn}dm/dv}}
.\end{equation}
We assumed that the energy that is produced by nuclear heating above this angle contributes to the latitudinal emission instantaneously, that is,~we set the latitudinal luminosity per unit velocity to (we assumed a uniform distribution of density in the latitudinal direction)
\begin{equation}
   \frac{dL_\mathrm{lat}}{dv}(v,t)= \frac{1}{2}\dot\epsilon(t) \frac{dm}{dv} \times \max\left(1-\frac{\theta_\mathrm{lat}(v)}{\theta_\mathrm{dyn}},1-\frac{\theta_\mathrm{d}(v,t)}{\theta_\mathrm{dyn}}\right)
,\end{equation}
where $\dot \epsilon(t)$ is the nuclear heating (energy per unit time, per unit mass) and  the factor $1/2$ accounts for the fact that we only considered the upper half of the ejecta, that is, region A in Fig.~\ref{fig:dyn_ej_emission_regions}.
The latitudinal surface area of a shell is
\begin{equation}
\frac{dS_{\mathrm{lat}}}{dv}(v,t)=\phi_\mathrm{dyn}v\,dv\,t^2
,\end{equation}
so that the effective temperature of the latitudinal annulus above the shell is
\begin{equation}
T_{\mathrm{BB,lat}}(v,t)=\left(\frac{dL_{\mathrm{lat}}/dv}{\sigma_\mathrm{SB}(dS_{\mathrm{lat}}/dv)}\right)^{1/4},
\end{equation}
where $\sigma_\mathrm{SB}$ is the Stefan-Boltzmann constant. 
As noted by \cite{Barnes2013}, when the temperature falls below the first ionisation temperature of lanthanides $T_\mathrm{La}\approx1000$ K, these elements recombine and the opacity drops sharply. The photosphere thus recedes inward, following the recombination front. During this process, the photospheric temperature remains constant at the recombination value. Thus we set
\begin{equation}\label{eq:Tbb}
T_{\mathrm{lat}}(v,t)=\max(T_{\mathrm{BB,lat}}(v,t),T_\mathrm{La}). 
\end{equation}

For the radially emitting part (region C in Fig.~\ref{fig:dyn_ej_emission_regions}), we used a similar approach, with a slight modification to account for the relative speed of the shell and the emitting surface: we assumed that all radiation escapes from the region for which the radial diffusion speed is higher than the local velocity (as in \citealt{Grossman2014}). This occurs beyond a `diffusion velocity' $v_\mathrm{d}$, which is obtained by solving the implicit equation
\begin{equation}
    t = \sqrt{\frac{\kappa_\mathrm{dyn}m_\mathrm{dyn}(>v_\mathrm{d})}{\theta_\mathrm{dyn}\phi_\mathrm{dyn}v_\mathrm{d}c}}
.\end{equation}

The luminosity in the radial direction is therefore given by $L_\mathrm{rad}(t) = \dot \epsilon \, m_\mathrm{rad}(>v_\mathrm{d}(t))$, where the mass $m_\mathrm{rad}(>v)$ is defined as
\begin{equation}
    m_\mathrm{rad}(>v) = \int_{v}^{v_\mathrm{max}} \frac{\theta_\mathrm{lat}(v)}{\theta_\mathrm{dyn}} \frac{dm}{dv}dv
,\end{equation}
and it represents the mass that moves faster than $v$ contained in region C of Fig.~\ref{fig:dyn_ej_emission_regions}.
The radially emitting surface is
\begin{equation}
S_\mathrm{rad}(t)\sim\phi_\mathrm{dyn}\theta_\mathrm{dyn}v_\mathrm{ph}^2 t^2 \, ,
\end{equation}
where the photospheric radius is again obtained by solving an implicit equation, 
\begin{equation}
    \tau = \frac{2}{3} = \frac{\kappa_\mathrm{dyn} m(>v_\mathrm{ph})}{ \theta_\mathrm{dyn}\phi_\mathrm{dyn}v_\mathrm{ph}^2 t^2}\, .
\end{equation}

\noindent The radial effective temperature, with the same assumptions as above, is then
\begin{equation}
T_\mathrm{rad}(t)=\max\left[\left(\frac{L_{\mathrm{rad}}(t)}{\sigma_\mathrm{SB}S_\mathrm{rad}(t)}\right)^{1/4},T_\mathrm{La}\right]. 
\end{equation}
\\ When we assume that the dynamical ejecta is geometrically thin, the projection factor for latitudinal emission for an observer at an angle $\theta_\mathrm{view}$ with respect to the polar axis is
\begin{equation}
f_\mathrm{lat}=\cos(\theta_\mathrm{view}).    
\end{equation}
The projection factor for radial emission is instead
\begin{equation}\begin{split}
f_\mathrm{rad}=&\,\pi \cos(\theta_\mathrm{view})\sin^2(\theta_\mathrm{dyn})+\\&+2\sin(\theta_\mathrm{view})[\theta_\mathrm{dyn}+\sin(\theta_\mathrm{dyn})\cos(\theta_\mathrm{dyn})].    
\end{split}\end{equation}
As a result, we computed the flux from the dynamical ejecta by integrating the latitudinal emission over the velocities (multiplied by 2 to account for the upper and lower edge) and from the radial surface, each multiplied by its projection factor, assuming blackbody spectra with the relevant temperatures. The mass of the dynamical ejecta $M_{\rm dyn}$ and their velocity $v_{\rm dyn}$ were derived using formulae from \cite{Kawaguchi2016}, as explained in section \ref{merger_result}.
Because of the tidal origin of the dynamical ejecta, weak interactions are not expected to change the matter composition significantly, and robust $r$-process nucleosynthesis always occurs inside it \citep{Roberts2017}. We therefore associate a high opacity $\kappa_{\rm dyn} = 15 \,{\rm cm^2~g^{-1}}$ with the dynamical ejecta. This is very different from the case where a supra- or hypermassive NS forms following a NSNS merger. In the latter case, high temperatures and strong neutrino irradiation can increase the electron fraction of matter expanding close
to the polar axis, inhibiting the production of lanthanides and lowering the photon opacity (\citealt{Fernandez2013}).

For the wind and secular ejecta, we considered parameters similar to those obtained in the analysis of the KN emission associated with GW170817. However, we modified a few parameters specific to the BHNS case to take the properties intrinsic to this type of binary into account.
The masses of the wind and viscous ejecta were calculated as 
fractions $\xi_{\rm w}=0.01$ and $\xi_{\rm s}=0.2$ of the disc mass \citep{Just2015,Fernandez2013,Metzger2014}.
The disc fraction of the wind ejecta in the BHNS case is notably smaller than for the NSNS case, where $\xi_{\rm w}$ could be a significant fraction of $\xi_{\rm s}$. Once gain, this is due to the absence of an intermediate supra- or hypermassive NS state that produces a stronger neutrino wind. We assumed the wind and secular ejecta opacities to be $1 \,{\rm cm^2~g^{-1}}$ and $5 \,{\rm cm^2~g^{-1}}$, respectively.

\section{Kilonova radio remnant}\label{knaft}
After producing the KN emission, the ejecta continue their expansion in the ISM. Because the faster ejecta move at supersonic speed, a shock forms. As slower ejecta cross the reverse shock, they contribute their energy to the shocked region. Based on energy and momentum conservation, the forward-shock radius $R$ can be related to the Lorentz factor $\Gamma$ of the shocked material \citep[e.g.][]{Hotokezaka2016} through
\begin{equation}
\Omega\frac{R^3}{3} m_{\rm p} n (c\beta\Gamma)^2 \sim E(>\beta),    
\end{equation}
where $\Omega$ is the solid angle subtended by the ejecta, $n$ is the ISM number density, $m_\mathrm{p}$ is the proton mass, and $E(>\beta)$ is the kinetic energy in the ejecta faster than $\beta c$, that is,
\begin{equation}
E(>\beta)=\int_{\beta c}^{v_\mathrm{max}} \left(\Gamma-1\right)\frac{dm}{dv}c^3\,d\beta
,\end{equation}
where $\Gamma=(1-\beta^2)^{-1/2}$ and $dm/dv$ is the distribution of ejecta mass in velocity space given in the preceding section. After the slowest ejecta have crossed the reverse shock, the expansion continues quasi-adiabatically, satisfying $\Gamma\beta\propto R^{-3/2}$. ISM electrons are accelerated at the shock and they emit by synchrotron radiation, mainly in the radio band. This emission component is sometimes referred to as a ``radio flare'' \citep{Nakar2011}, but given its very slow evolution (typically on a timescale of several years) and because it is essentially the same as the shock-related component of a supernova radio remnant, we prefer the nomenclature ``kilonova radio remnant''. 
Predictions for this emission component have been made previously (e.g. in \citealt{Hotokezaka2015}) for both the NSNS and BHNS case. Its peak flux density, which can in principle reach the mJy level on several-year timescales, is highly uncertain, however, because it depends strongly on several of the assumed parameters (e.g. the ISM density).

As in \cite{Nakar2011}, we modelled the synchrotron emission from the shocked material following a treatment similar to GRB afterglows \citep{Sari1998}: electrons behind the forward shock were assumed to be accelerated (e.g. through the Fermi process) into a power-law energy distribution of index $p$; their total energy density was assumed to be a fraction $\epsilon_\mathrm{e}$ of the energy density behind the shock, which is set by the shock jump conditions \citep{Blandford1976}; similarly, the magnetic field behind the shock was assumed to be amplified by small-scale instabilities to an energy density equal to a fraction $\epsilon_\mathrm{B}$ of the total energy density. We modelled synchrotron self-absorption following \cite{Panaitescu2000}. We only considered dynamical ejecta (for which we set $\Omega=\theta_\mathrm{dyn}\phi_\mathrm{dyn}$), as the disc winds and viscous ejecta are much slower, which results in a much later deceleration. For our assumed parameters, their radio remnant becomes relevant only later than $10^4\,\mathrm{days}$.

\section{Relativistic jet launch}\label{jet}

When a disc remains after the BHNS merger, its accretion onto the final BH can cause the launch of a relativistic jet via the Blandford-Znajek mechanism \citep{Blandford1977,Komissarov2001}. The luminosity that can be extracted by this process is \citep{Tchekhovskoy2010}

\begin{equation}
\label{eq:Lbz}
L_{\rm BZ}\propto \frac{G^2}{c^3} M_\mathrm{BH}^2 B^2 \Omega_{\rm H}^2 f(\Omega_\mathrm{H}),
\end{equation}
where $B$ is the magnetic field at the BH event horizon, $0\leq \Omega_{\rm H}\leq 1/2$ is the dimensionless angular frequency at the horizon,
\begin{equation}
\Omega_{\mathrm {H}}=\frac{\chi_\mathrm{BH}}{2(1+\sqrt{1-\chi_\mathrm{BH}^2})},\end{equation}
and $f(\Omega_\mathrm{H}) = 1+1.38\Omega_\mathrm{H}^2-9.2\Omega_\mathrm{H}^4$ is a high-spin correction. 
\\In this formula $\chi_{\rm BH}$ is the spin parameter of the final BH: we computed this quantity using eq. 11 from \cite{Pannarale2013}.

Under the assumption that the magnetic field is amplified by Kelvin-Helmholtz and magneto-rotational instabilities (MRI) in the post-merger phase and reaches equipartition with the disc energy density \citep{Giacomazzo2015}, we have that approximately

\begin{equation}
B^2 \propto \frac{c^5}{G^2} \dot{M}M_\mathrm{BH}^{-2},
\end{equation}
where $\dot{M}$ is the mass accretion rate onto the BH, and thus
\begin{equation}
\label{eq:Lbz_eff}
L_{\rm BZ}\propto \dot{M} c^2 \Omega_{\rm H}^2 f(\Omega_\mathrm{H}).
\end{equation}
This scaling has been found to be in agreement with results of general relativity magneto--hydrodynamic (GRMHD) simulations of compact object mergers that launch a jet \citep{Shapiro2017}. 

\begin{figure}
    \centering
    \includegraphics[width=\columnwidth]{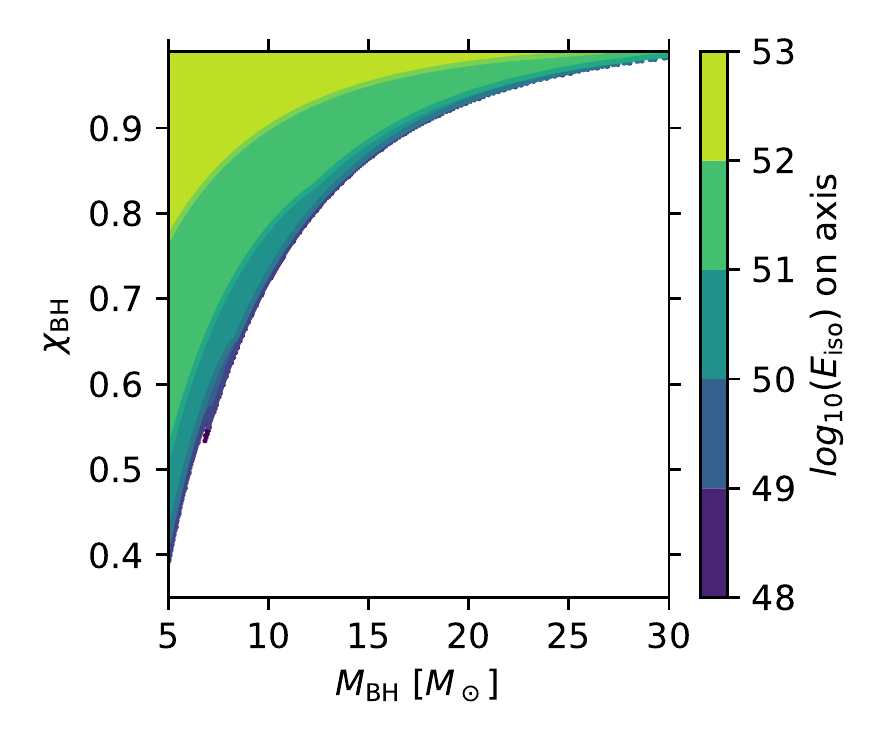}
    \caption{Isotropic equivalent energy of the jet core to be radiated as GRB prompt emission (as seen from an on-axis observer) as a function of the BH mass and spin. }
    \label{fig:E_onaxis}
\end{figure}
After the launch, the jet may loose some energy upon interaction with the ambient medium (i.e.~the other merger ejecta). We assumed the jet to be launched in the polar direction (perpendicular to the accretion disc). Along this direction, the density of the ejecta  is likely very low. In a BHNS merger the dynamical ejecta produced by the tidal disruption are indeed launched close to the equatorial plane (e.g. \citealt{Kawaguchi2016} opening angles $\theta_{\rm{dyn}}\leq 22^\circ$). No shocks, as in the NSNS case (when the two stars collide), are expected in the BHNS case. Shocks would produce a much more isotropic ejection of matter. 

Because of the centrifugal force in the disc co-rotating frame, the viscous ejecta are distributed approximately as $\sin^2\theta$ \citep{Perego2017}. Thus, only a small fraction of their mass is contained in the polar region. 

Finally, the wind ejecta, which represent the only outflow preferentially emitted along the pole, contribute only very little mass, as explained in the previous section. Therefore, we assumed that the jet overcomes the ejecta, which spends only a negligible fraction of its energy, without consequences on its structure. Its kinetic energy is therefore $E_\mathrm{K,jet}=L_{\rm BZ}\times t_{\rm acc}$ where $t_{\rm acc}$ is the disc accretion time. Because $t_\mathrm{acc}=(1-\xi_\mathrm{w}-\xi_\mathrm{s})M_\mathrm{disc}/\dot{M}$ (where the factor in parentheses accounts for the disc mass lost in winds, and thus not accreted), we have
\begin{equation}
    E_\mathrm{K,jet} = \epsilon (1-\xi_\mathrm{w}-\xi_\mathrm{s})M_\mathrm{disc}c^2\,\Omega_\mathrm{H}^2 f(\Omega_\mathrm{H}).
    \label{eq:jet_kinetic_energy}
\end{equation}
The dimensionless proportionality constant $\epsilon$ depends on the ratio of magnetic field energy density to disc pressure at saturation \citep{Hawley2015}, on the large-scale magnetic field geometry, and on the disc aspect ratio \citep{Tchekhovskoy2010}, but there are indications \citep{Shapiro2017} that it is the same across BHNS mergers. In order to set it to a definite value, we determined its upper extremum. First, we note that the maximum disc mass cannot exceed the total NS baryonic mass, $M_\mathrm{disc}\lesssim 2 \,\mathrm{M_\odot}$, and the spin-dependent factor $\Omega_\mathrm{H}^2 f(\Omega_\mathrm{H})$ cannot exceed $0.2$. The most energetic short GRB observed so far had $E_\mathrm{\gamma,iso}\sim 7.4\times 10^{52}\mathrm{erg}$ \citep[GRB 090510, ][]{Davanzo2014}: assuming a typical 10\% conversion efficiency of kinetic to gamma-ray energy and using a jet half-opening angle of $5\,\mathrm{deg}$ \citep[the typical measured half-opening angle of SGRBs, see][]{Fong2015}, we have that this corresponds to $E_\mathrm{K,jet}\sim 3\times 10^{51}\mathrm{erg}$. Based on these considerations, we set $\epsilon=0.015$, which sets the maximum possible jet energy release to $E_\mathrm{K,jet,max} \approx 10^{52}\mathrm{erg}$.

\begin{figure}
    \centering
    \includegraphics[width=\columnwidth]{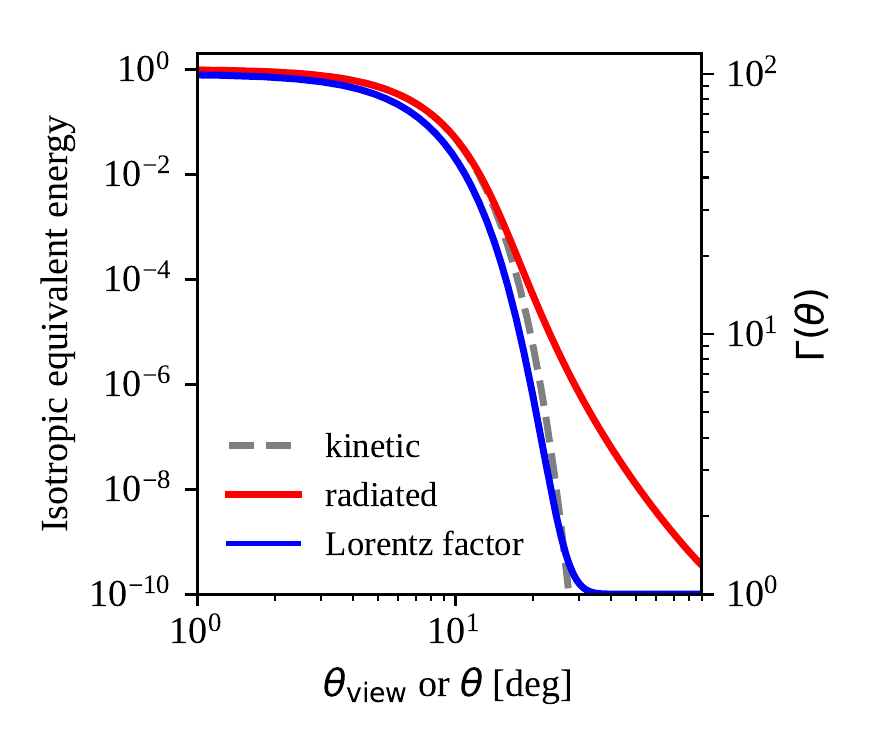}
    \caption{Jet structure functions $E_\mathrm{K,iso}=4\pi dE/d\Omega(\theta)$ (dashed grey line, normalised to the value at the jet axis) and $\Gamma(\theta)$ (solid blue line, values shown on the right vertical axis). The solid red line shows the radiated isotropic-equivalent energy $E_\mathrm{iso}(\theta_{\rm view})$ normalised to the value measured by an on-axis observer as a function of the viewing angle.}
    \label{fig:Eiso_thv}
\end{figure}
\subsection*{Jet structure}
As shown by~\cite{Kath2018}, for instance, a jet launched by magnetohydrodynamic energy extraction from a spinning BH naturally develops an angular distribution of Lorentz factor $\Gamma$ and kinetic energy per solid angle. Both quantities decrease approximately exponentially with the angular distance from the jet axis.

\begin{figure*}
    \centering
    \includegraphics[width=\textwidth]{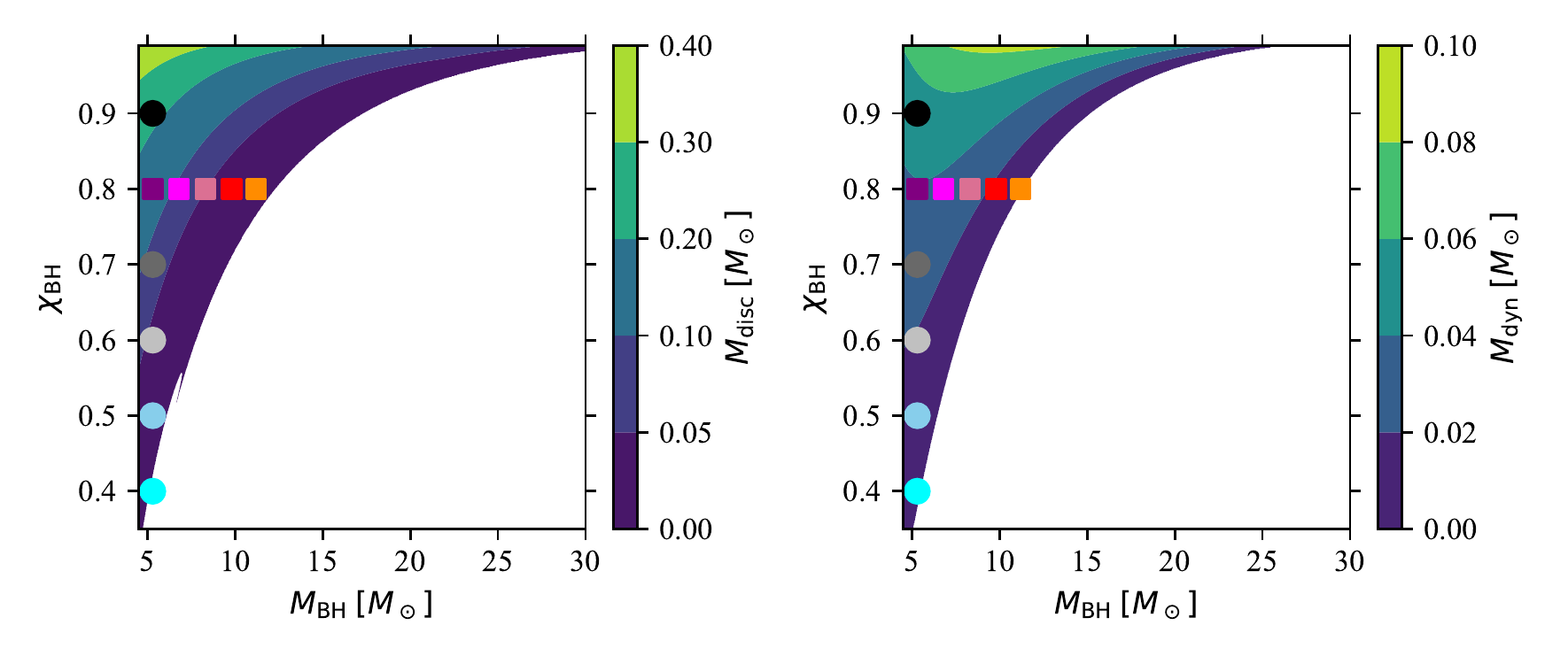}
    \caption{Accretion disc (left) and dynamical ejecta (right) masses in the $M_\textrm{BH}-\chi_\textrm{BH}$ plane. We show two parameter sets: squares have constant spin and different masses, and circles have constant mass and different spins. These sets are used to produce the example light curves shown in Fig.\ref{fig:example_lightcurves}.}
    \label{fig:masse}
\end{figure*}
We assumed the following angular distributions, inspired by those found by \cite{Kath2018}:
\begin{equation}\begin{split}
&\frac{dE}{d\Omega}(\theta)=E_\mathrm{c}e^{-(\theta/\theta_\mathrm{c,E})^2};
\\&\Gamma(\theta)=(\Gamma_\mathrm{c}-1)e^{-(\theta/\theta_\mathrm{c,\Gamma})^2}+1;
\end{split}\end{equation}
where we set $\Gamma_\mathrm{c}=100$, $\theta_\mathrm{c,E}=0.1$ rad, $\theta_\mathrm{c,\Gamma}=0.2$ rad, and $E_\mathrm{c}= E_\mathrm{K,jet}/\pi\theta_\mathrm{c,E}^2$. 

This structure, shown in Figure ~\ref{fig:Eiso_thv}, represents an educated guess that will be compared with observations of real sources in the future. Given the likely absence of substantial collimation by the ambient material, the jet structure in these types of systems should keep some memory of the launching region (e.g.~the magnetic field configuration). If, speculatively, the launch conditions were the same across different systems, these jets could then feature a quasi-universal structure, that is,~they could differ only by a small scatter in their properties.

\section{Gamma-ray burst prompt emission}\label{prompt}
\begin{figure*}
    \centering
    \includegraphics[width=\textwidth]{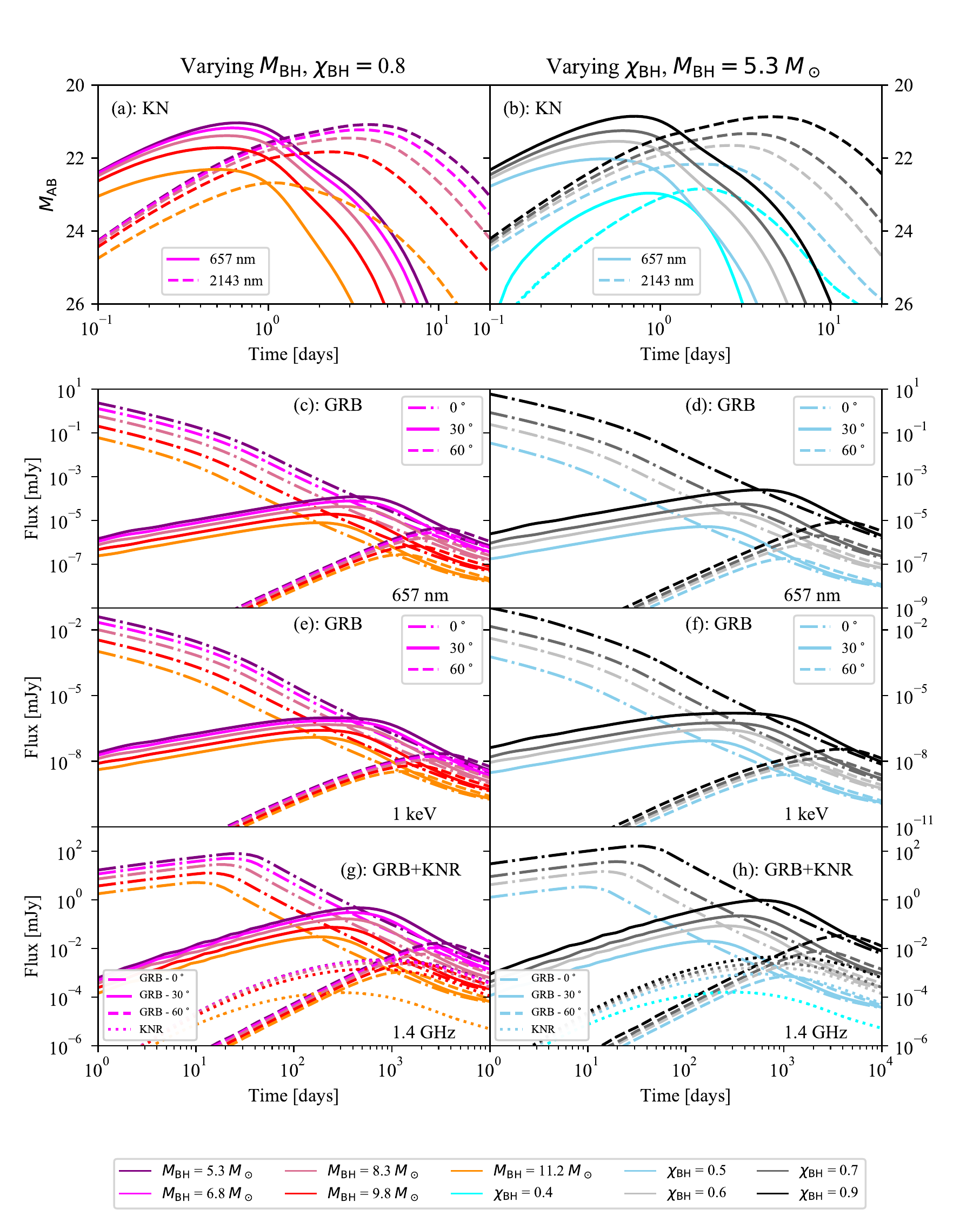}
    \caption{Example light curves for constant $\chi_\mathrm{BH}$ - varying $M_\mathrm{BH}$ (left columns) and constant $M_\mathrm{BH}$ - varying $\chi_\mathrm{BH}$ (right columns). Panels a-b: KN r-band ($657$ nm, filled lines) and K-band ($2143$ nm, dashed lines) light curves. Panels c-h: GRB afterglow optical (panels c-d), X-ray  (panels e-f), and radio (panels g-h) light curves for three viewing angles ($0^\circ$ dot-dashed lines, $30^\circ$ filled lines, and  $60^\circ$ dashed lines). The KNR (dotted lines) is also shown in panels g-h.}
    \label{fig:example_lightcurves}
\end{figure*}
Following standard practice, we assumed that a fraction   $\eta=10\%$ of the kinetic energy in the jet is dissipated (\eg by internal shocks or magnetic reconnection) and radiated. The isotropic equivalent energy in radiation, as seen by an observer at a viewing angle $\theta_\mathrm{v}$, is then given by \citep{Salafia2015}
\begin{equation}
    E_\mathrm{iso}(\tv)=\eta \int \frac{\delta^3}{\Gamma}\frac{dE}{d\Omega}d\Omega
.\end{equation}
In Figure~\ref{fig:E_onaxis} we show the isotropic equivalent radiated energy (as measured by an on-axis observer) as a function of the BH intrinsic properties. As anticipated before, the jet will be launched only if an accretion disc is formed, thus no values are given for parameters that result in a direct plunge of the NS onto the BH. We note that the energy range obtained by our modelling corresponds to the observed range of energies of short GRBs \citep[e.g.][]{Davanzo2014}. Figure~\ref{fig:Eiso_thv} shows the dependence of $E_\mathrm{iso}$ on the viewing angle for the assumed jet structure, along with the assumed structure functions $E_\mathrm{K,iso}=4\pi dE/d\Omega(\theta)$ and $\Gamma(\theta)$.

\section{Gamma-ray burst afterglow}\label{GRBafterglow}
After producing the prompt emission, the jet continues to expand into the ISM. As soon as a sufficient amount of ISM matter is swept away, a strong forward shock forms, which gives rise to the jet afterglow. We computed the forward shock dynamics (neglecting lateral spreading) and its synchrotron emission using an updated version of the model employed in \citet{D'Avanzo2018} and \citet{Ghirlanda2018} (a detailed description of the model will be given in Salafia et al. 2019, in preparation). The synchrotron emission parameters are the same as for the KNR described above.

To produce the example light curves shown in Figure \ref{fig:example_lightcurves}, we assumed a constant ambient medium density $n=10^{-3}$ cm$^{-3}$. This value is consistent with the few short GRBs whose afterglow emission has been modelled \citep{Fong2015} and with estimates of this parameter in the NSNS event 170817 \citep[e.g.][]{Ghirlanda2018}. Moreover, it is consistent with the expectations for a low-density ambient medium such as the site where binaries might merge as a result of supernova kicks. The fractions of shock energy carried by electrons and magnetic field are assumed to be $\epsilon_{\rm e}=0.1$ and $\epsilon_{\rm B}=0.01$, respectively. Although it is hardly constrained from  afterglow observations, the value $\epsilon_{\rm e} = 0.1$ has been shown to be typical based on the analysis of the radio to GeV emission energy ratio in long GRBs \citep{Beniamini2017,Nava2014}. The value of $\epsilon_{\rm B}$ is less well constrained and can be distributed between 10$^{-4}$ and 10$^{-1}$ \citep[e.g.][]{Granot2014,Santana2014,Zhang2015,Beniamini2016}. Finally, we assumed a non-thermal energy distribution of shock-accelerated electrons with slope parameter $p=2.3$, as expected based on particle-in-cell simulations of Fermi acceleration in mildly magnetised relativistic shocks (e.g. \citealt{Sironi2013}).

\section{Example light curves: dependence on BH spin and mass}\label{lightcurves}
In this section we describe some key dependencies of the multi-band counterpart light curves on the main intrinsic parameters of the system: the black hole mass and spin. For this purpose, we selected a set of reference points in the $M_{\rm BH}-\chi_{\rm BH}$ plane, shown in Figure~\ref{fig:masse}. The set consist of combinations with the same BH mass and different BH spins (circles in Figure~\ref{fig:masse}) and of others with the same BH spin and different BH masses (squares). 
 
In panels a and b of Figure~\ref{fig:example_lightcurves} we show the KN light curves for the two sets of points, in the r band ($657$ nm, filled lines) and K band ($2143$ nm, dashed lines). It is apparent that the lower the BH mass and the higher the spin, the brighter the KN (the more massive the ejecta). In addition to the brightness, BH mass and spin also affect the shape of the KN light curve: for lower BH masses and higher spins, the peaks shift at later times.

The case with $M_\mathrm{BH}=5.3$ $M_\odot$ and $\chi_\mathrm{BH}=0.4$, denoted with blue lines, is interesting because only dynamical ejecta are present without a disc. As shown in Figure~\ref{fig:masse}, the corresponding KN is much dimmer than the others (being produced only by one ejecta component) and there is no GRB afterglow (no relativistic jet is produced because there is no accretion disc). This is clearly a limiting case because in reality, it seems unlikely that the tidal disruption of the NS can lead to the production of only unbound material. A small mismatch between the two fitting formulae for the disc and ejecta masses in this region of the parameter space is the most reasonable explanation for this particular case.

In panels c-h of Figure~\ref{fig:example_lightcurves} we show the GRB afterglow light curves for the two sets of parameters for three different viewing angles. Panels c-d show the optical emission in the r band ($657$ nm), panels e-f plot the X-ray emission ($1$ KeV), and panels g-h show the radio emission ($1.4$ GHz).
In panels g-h we also show the KNR (dotted lines). In both cases, brighter emission corresponds to lower BH mass/higher BH spin (more massive ejecta).

Figure~\ref{fig:example_lightcurves} shows that the light curves 
from BHNS are highly degenerate although their time behaviour is closely correlated with the dynamics of the BHNS debris. For the single EM multi-band light curve, it is impossible to infer the intrinsic parameters at the source, and in particular, the $M_{\rm BH},\chi_{\rm BH}$ degeneracy that emerges from the figure. The concordant  analysis from the GW signal and EM light curve together may help to brake this degeneracy, however. The BH and NS masses together with the luminosity distance $d_\mathrm{L}$ can be inferred from the GW signal. The identification of the host galaxy from the EM counterpart provides the redshift of the source, thus narrowing the uncertainties in the parameter estimation of  $M_\mathrm{BH}$, $M_\mathrm{NS}.$  Under these conditions, the light curve carries valuable information on the BH spin that can be inferred from the EM observation.

\section{Test case: constraining the BH spin}\label{testcase}

We considered a BHNS merger with parameters in the source frame  $M_\textrm{BH}=6\,M_\odot$, $\chi_\textrm{BH}=0.8$, $M_\textrm{NS}=1.4\,M_\odot$, and $\Lambda_{\rm NS}=330$. These parameters correspond to a chirp mass $M_\mathrm{c}\approx 2.4\msun$. As stated in section \ref{parameters}, we assumed $d_\textrm{L}=230$ Mpc, $\iota_\textrm{tilt}=0$ rad and $\theta_\textrm{view}=30^\circ$. According to the fitting formulae described in \S\ref{merger_result}, upon merger, this binary would produce $0.038\msun$ of dynamical ejecta and an accretion disc with a mass of $0.114\msun$. Likewise, according to Eq.~\ref{eq:jet_kinetic_energy}, the merger remnant would produce a jet with a total kinetic energy $E_\mathrm{K,jet}\sim 1.6\times 10^{50}\,\mathrm{erg}$ and an on-axis isotropic-equivalent energy $E_\mathrm{iso,on-axis}\sim 6.4\times 10^{51}\,\mathrm{erg}$ (assuming a 10\% efficiency in kinetic-to-radiated energy conversion). Fig.~\ref{fig:Eiso_thv} shows that a $30^\circ$ off-axis observer would see this energy reduced by a factor $\sim 10^{-6}$, which would make the prompt emission of this jet essentially undetectable at $230\,\mathrm{Mpc}$ with current facilities.

\begin{figure*}
    \centering
    \includegraphics{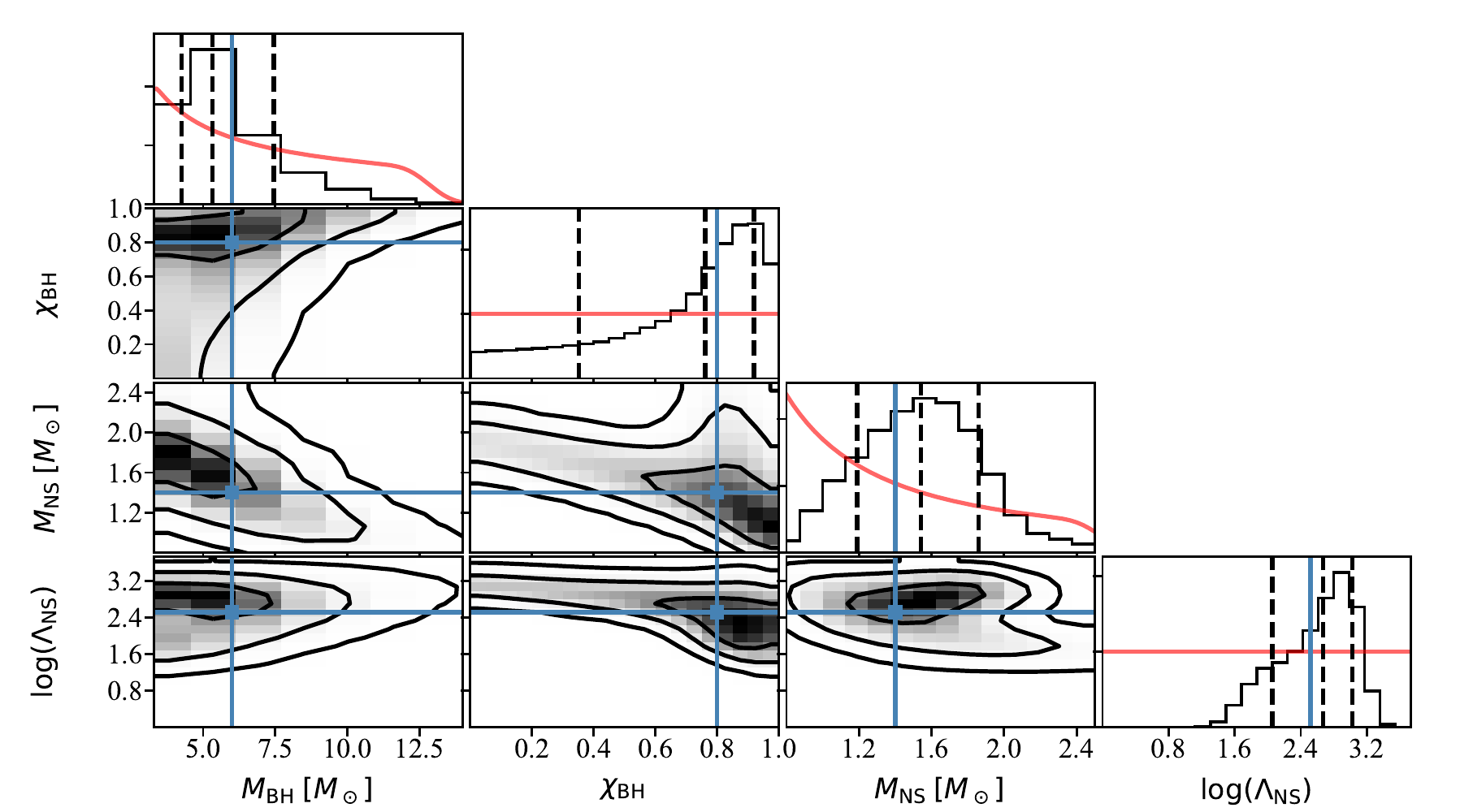}
    \caption{Marginalised posterior distributions for $M_\textrm{BH}$, $\chi_\textrm{BH}$, $M_\textrm{NS}$, log($\Lambda_\mathrm{NS}$), and joint posteriors for all pairs of these parameters. Black dashed lines in the plots on the diagonal represent the value corresponding to the 16th (left line), 50th (central line), and 84th (right line) percentiles. Red lines show the prior distributions. Black lines in the joint posterior plots represent one-, two- and three-sigma contours, while black dots show single samples in the region outside the three-sigma contour. Blue lines indicate the original values from which the mock data set have been produced.}
    \label{fig:corner}
\end{figure*}

\begin{figure*}
    \centering
    \includegraphics{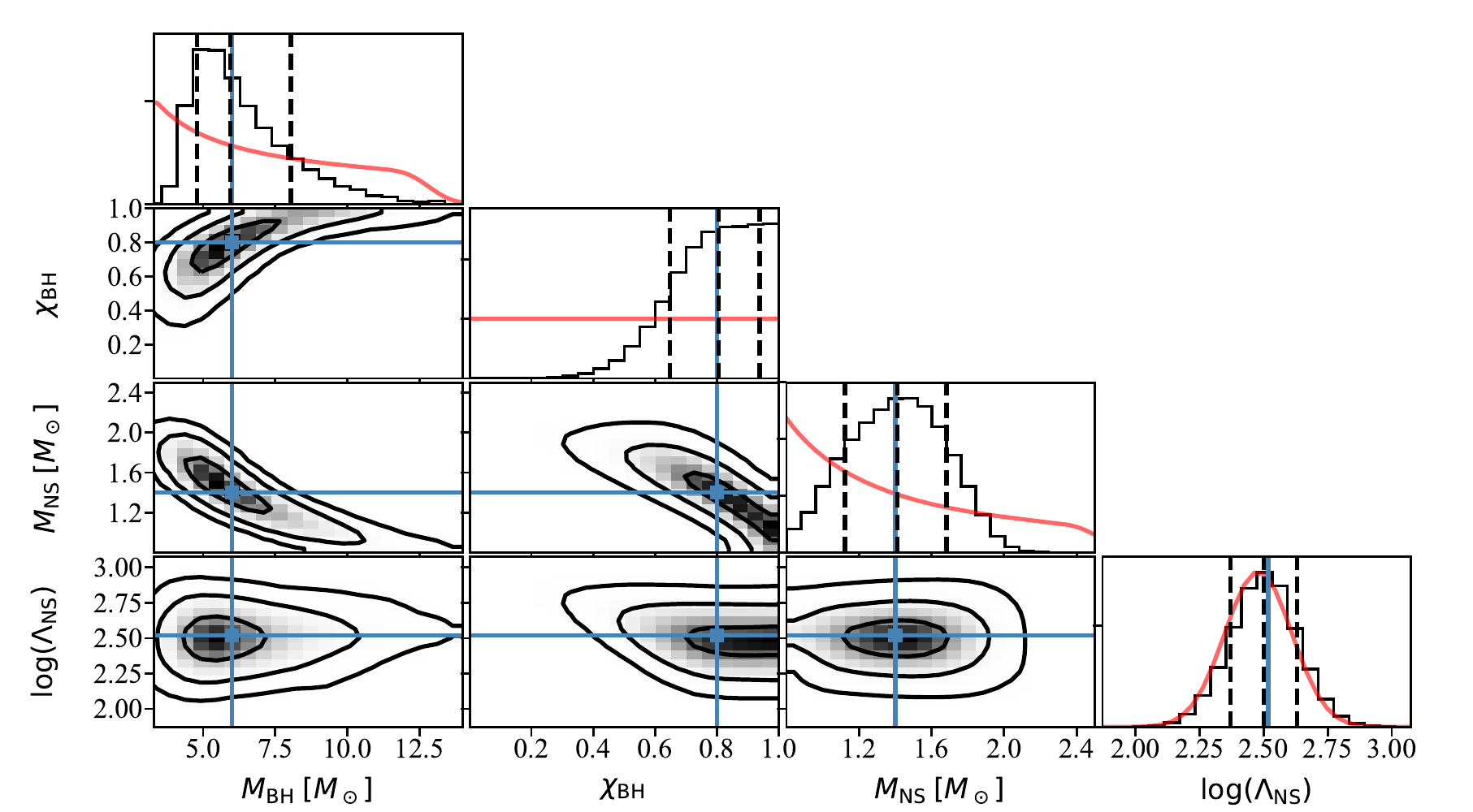}
    \caption{Same as Figure \ref{fig:corner}, assuming a log-normal prior for $\Lambda_\mathrm{NS}$.}
    \label{fig:corner_2}
\end{figure*}

Furthermore, we expect the KNR and the GRB afterglow (for this viewing angle) to peak $\text{about some}$ hundreds of days after the merger (see Figure~\ref{fig:example_lightcurves}). In this test case, we therefore considered only the KN (visible from $\text{approximately}$ hours to $\text{some}$ tens of days) as the EM counterpart to be employed in the analysis.

We considered two wavelengths: $657$ nm (r band, optical) and $2143$ nm (K band, infrared). We created the mock data points by selecting evenly spaced times from $0.1$ days to $30$ days and assuming a constant error on magnitudes ($0.2$ mag for r band and $0.3$ mag for K band). We imposed limiting observation magnitudes of $28$ for the r band and $24$ for the K band.
We then performed a Markov chain Monte Carlo (MCMC) analysis on our mock data set to constrain the BH spin, adopting the \texttt{emcee} sampler \citep{emcee}. 

The free parameters in our MCMC were the BH mass $M_\mathrm{BH}$, the BH spin $\chi_\mathrm{BH}$, the NS mass $M_\textrm{NS}$ , and the NS dimensionless quadrupolar tidal deformability $\Lambda_\textrm{NS}$.
We assumed a flat prior on $\chi_\textrm{BH}$ in the $[0,0.99]$ interval and a log-flat prior on $\Lambda_\mathrm{NS}\in[10,10^4]$. Because we wished to simulate a multi-messenger analysis, we also included (at least in a simplified form) the information from the GW signal. We did this in a simple way by assuming that the GW analysis yields a Gaussian posterior on the chirp mass only. We converted this posterior into a two-dimensional prior on the BH and NS masses for use in our EM analysis, as computed in \S\ref{mbh_gw}.

In Figure ~\ref{fig:corner} we show the resulting marginalised posterior distributions for the four parameters and the joint posterior distributions of parameter pairs. Blue lines and squares indicate the original values from which the mock data have been produced. Red lines show the priors.

\begin{figure}
    \centering
    \includegraphics[width=\columnwidth]{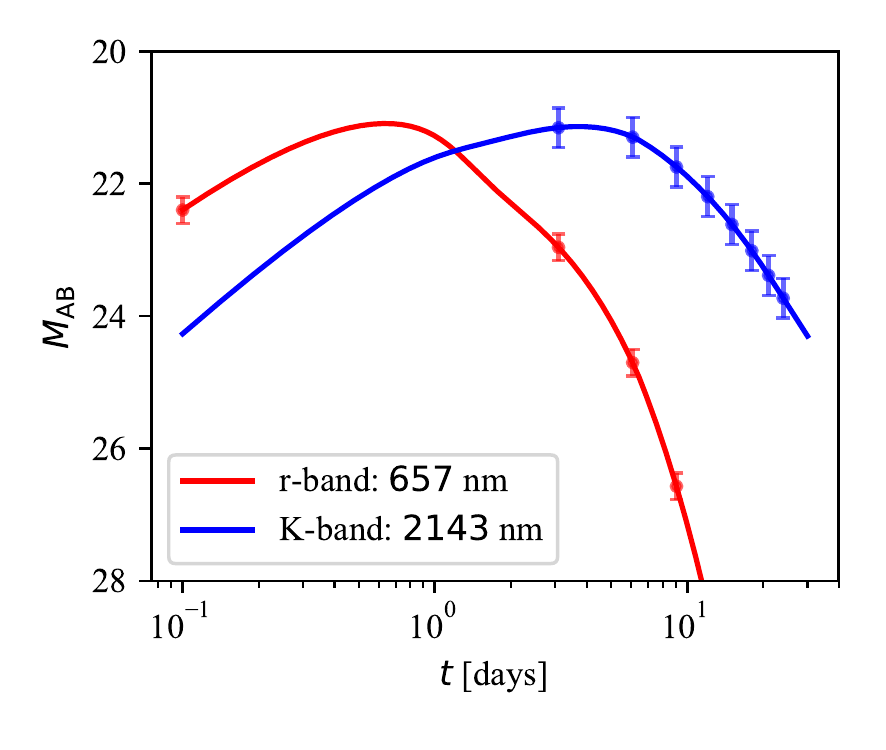}
    \caption{Mock photometric data in r band (blue) and K band (red) with errors (1$\sigma$) and light curves corresponding to best-fit values.} 
    \label{fig:fit_lc}
\end{figure}

We calculated the best-fitting parameter values following the method described in \cite{Ghirlanda2019}. The fit results are presented in Table \ref{tab:fitresult} (left column). In Fig.~\ref{fig:fit_lc} we show the mock photometric data with errors and the model light curves that correspond to the best-fit values.

The parameter estimates are consistent with the input estimates, demonstrating that the light curves encode information about the progenitor binary, through their dependence on the ejecta properties. The residual bias in the best-fit values is essentially due to the broad uninformative prior assumed for $\Lambda_\mathrm{NS}$. In order to show this, we performed as a proof of concept a second parameter estimation using a log-normal prior on $\Lambda_\mathrm{NS}$, centred at $\Lambda_\mathrm{NS}=330$, with $\sigma=0.3$. 

The fit results are presented in Table \ref{tab:fitresult} (right column). In Figure \ref{fig:corner_2} we present the ‘corner' plot (same legend as Figure \ref{fig:corner}). It is clear that in this case the best-fitting values are much closer to the 'true' values. The BH spin is constrained with an unprecedented precision.

\begin{table}
    \caption{Best-fit parameter values obtained using two possible priors on $\Lambda_\mathrm{NS}$.}
    \centering
    \begin{tabular}{lcc}
    \toprule
         & log-flat prior on $\Lambda_\mathrm{NS}$ & log-normal prior on $\Lambda_\mathrm{NS}$\\
        \addlinespace[0.08cm]
        \midrule
        $M_\mathrm{BH}$ [$\msun$] & $5.6^{+1.9}_{-1.3}$ & $6.1^{+2.0}_{-1.3}$ \\
        \addlinespace[0.1cm]
        $\chi_\mathrm{BH}$ & $0.7^{+0.2}_{-0.3}$ & $0.8^{+0.1}_{-0.2}$ \\ 
        \addlinespace[0.1cm]
        $M_\mathrm{NS}$ [$\msun$] & $1.5^{+0.4}_{-0.3}$ & $1.4^{+0.3}_{-0.3}$ \\
        \addlinespace[0.1cm]
        $\log(\Lambda_\mathrm{NS})$ & $2.7^{+0.3}_{-0.6}$ & $2.5^{+0.2}_{-0.1}$ \\
        \midrule
    \end{tabular}
    \label{tab:fitresult}
\end{table}

\section{Conclusion}
The BHNS coalescences can be exquisite multi-messenger sources. No GW signal has been observed so far from this family of sources, but EM transient emission from this type of merger may have already left its imprint on the light curve of some observed GRBs. 

We built a composite model to describe the complex EM signal that accompanies the tidal (partial) disruption of an NS during its inspiral and plunge across the horizon of a stellar BH. It is known that mass shedding, which is required to produce an EM counterpart, can only emerge over a limited range of mass ratios, BH spins, and degrees of NS tidal deformability. We estimated the amount of this mass as a function of these quantities using physically motivated numerical-relativity-informed fitting formulae from the literature (\S\ref{merger_result}). 

Our composite model includes KN emission from the dynamical ejecta and disc winds, accounting for their expected anisotropies, in both their geometry and composition (\S\ref{Kilonova}). Additionally, it predicts the late-time emission from the radio remnant that is expected to be associated with the deceleration of the dynamical ejecta into the ISM (\S\ref{knaft}). It also includes the prompt (\S\ref{prompt}) and afterglow (\S\ref{GRBafterglow}) emission from the relativistic jet that might be launched by the merger remnant, accounting for its anisotropic properties (its energy and Lorentz factor angular distribution, \S\ref{jet}). 

We presented a suite of light curves obtained by varying the BH mass and spin (Figure~\ref{fig:example_lightcurves}), one at a time, to show the variety in the prospected emission from these coalescences. These light curves show a high degree of degeneracy that is produced by different parameter combinations. It is thus impossible to infer the intrinsic parameters of the source using only the EM multi-band light curves. However, by joining the information from GW and EM signal analysis, it is possible to break this degeneracy. By constraining the BH and NS masses from the GW signal and the redshift from the EM counterpart, we can indeed extract valuable information on the BH spin from light curves.
\\As a proof of concept, we proposed an example of joint multi-messenger analysis. In order to represent the information that comes from the GW analysis in a simple way, we assumed it to be encoded into a simple Gaussian posterior on the chirp mass. We then used it as a prior for the EM analysis. For simplicity, we considered only the KN as observed EM counterpart, and we limited ourselves to only two wavelengths. Our results show that the joint analysis results in a constraint on the BH spin, even in our very conservative setting (which may be considered as representative of a GW detection with a very low signal-to-noise ratio) in which the GW signal provides only information on the chirp mass. We will explore how including the other emission components, considering more wavelengths, and taking into account the whole information from the GW analysis can improve the constraints on the intrinsic binary parameters.

\begin{acknowledgements}
We thank S. Bernuzzi for useful comments and suggestions during the preparation of the first draft. O.~S. acknowledges the Italian Ministry for University and Research (MIUR) for funding through project grant 1.05.06.13. The authors acknowledge support from INFN, under the Virgo-Prometeo initiative.
\end{acknowledgements}

\footnotesize{
\bibliographystyle{aa}
\bibliography{references}
}

\section*{Appendix A: Mass posterior from the GW signal}\label{mbh_gw}

\renewcommand\thefigure{A.1} 

Observation of GW from a compact binary inspiral (BHNS in our case) provides (at least) a measure of the chirp mass
\begin{equation}\label{eq:chirp}
M_\mathrm{c}=\frac{(M_{\rm BH}M_{\rm NS})^{3/5}}{(M_{\rm BH}+M_{\rm NS})^{1/5}}
.\end{equation}
The uncertainty on $M_\mathrm{c}$ is broader for higher chirp masses because more massive systems emit in the detector band for a shorter time prior to merger. By the Bayes theorem, the probability for $M_{\rm BH}$ and $M_{\rm NS}$ given a measured $M_c$ is
\begin{equation}\label{eq:P(MBH,MNS)}
P(M_{\rm BH},M_{\rm NS}|M_\mathrm{c})=\frac{P(M_\mathrm{c}|M_{\rm BH},M_{\rm NS})P(M_{\rm BH})P(M_{\rm NS})}{P(M_\mathrm{c})}
.\end{equation}

We assumed that the uncertainty on the measured chirp mass  is represented by a Gaussian centred around the true value, 
\begin{equation}\label{eq:P(MC)}
P(M_\mathrm{c}|M_{\rm BH},M_{\rm NS})\propto \exp{\left[-\frac{1}{2}\left(\frac{M_\mathrm{c}-\frac{(M_{\rm BH}M_{\rm NS})^{3/5}}{(M_{\rm BH}+M_{\rm NS})^{1/5}}}{\sigma_\mathrm{c}}\right)^2\right]}. 
\end{equation}

For BHBH mergers detected by Advanced LIGO\& Virgo during O1 and O2 the relative error $e_{\mathrm{M}_\mathrm{c}} \in [2-20\%]$, while for the GW170817 NSNS merger $e_{\mathrm{M}_\mathrm{c}}\approx0.1\%$ \citep{catalogo_GW}. A BHNS merger is an intermediate case between the two, therefore we conservatively assumed for our example case $e_{\mathrm{M}_\mathrm{c}}\approx 2$, which sets the $\sigma_\mathrm{M_c}$ parameter above.

\begin{figure}
    \centering
    \includegraphics[width=\columnwidth]{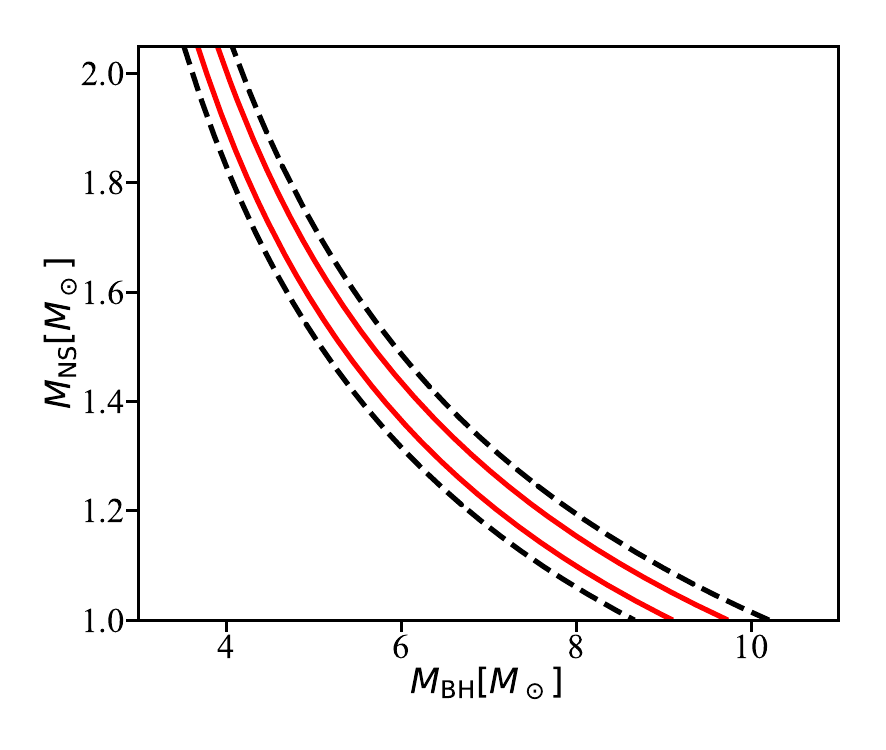}
    \caption{Example of bidimensional posterior distribution for BH and NS masses from a simulated  GW analysis. Solid (red)  and black (dashed) lines represent the $50\%$ and $90\%$ confidence levels, respectively.}
    \label{fig:mbh_mns_post_gw}
\end{figure}

We therefore defined the GW analysis two-dimensional \textit{\textup{posterior}} on the BH and NS masses by Eqs.~\ref{eq:P(MBH,MNS)} and \ref{eq:P(MC)}, and we used this as a\textup{ prior} for the EM analysis. For our example BHNS merger with $M_{\rm BH}=6\, {\rm M_\odot}$ and $M_{\rm NS}=1.4\,{\rm M_\odot}$, the chirp mass (Eq.~\ref{eq:chirp}) equals $2.402\,{\rm M_\odot}$. Considering a measured chirp mass $M_\mathrm{c}=2.403\pm0.05\,{\rm M_\odot}$, we obtained the posteriors of the  BHNS masses shown in Figure \ref{fig:mbh_mns_post_gw}. This curve represents all the combinations of $M_{\rm BH}$ and $M_{\rm NS}$ that give a chirp mass consistent with the measured mass.

\end{document}